\documentclass[preprint,aps,nofootinbib,showkeys,showpacs,12pt]{revtex4}
\usepackage{amsmath,amssymb,amsfonts}
\usepackage{graphicx}% Include figure files
\usepackage{hhline}

\newcommand{\be}{\begin{eqnarray}}
\newcommand{\ee}{\end{eqnarray}}
\newcommand{\beq}{\begin{eqnarray}}
\newcommand{\eeq}{\end{eqnarray}}

\newcommand{\Z}{\mathbb{Z}}

\newcommand{\bc}{\begin{center}}
\newcommand{\ec}{\end{center}}

\newcommand{\Tcc}{T^\chi_c}

\newcommand{\Ep}{ E_i^{(+)}}
\newcommand{\Em}{ E_i^{(-)}}

\newcommand{\expm}{\mathrm{e}^{-\beta \Em}}
\newcommand{\expmm}{\mathrm{e}^{-2\beta \Em}}
\newcommand{\expmmm}{\mathrm{e}^{-3\beta \Em}}

\newcommand{\expp}{\mathrm{e}^{-\beta\Ep}}
\newcommand{\exppp}{\mathrm{e}^{-2\beta\Ep}}
\newcommand{\expppp}{\mathrm{e}^{-3\beta\Ep}}

\begin{document}
\title{\bf\large Scalar-pseudoscalar meson behavior and restoration of symmetries
in SU(3) PNJL model}
\author{\bf P. Costa}
\email{pcosta@teor.fis.uc.pt}
\affiliation{Departamento de F\'{\i}sica, Universidade de
Coimbra, P-3004-516 Coimbra}
\affiliation{E.S.T.G., Instituto Polit\'ecnico de Leiria,
Morro do Lena-Alto do Vieiro, 2411-901 Leiria, Portugal}
\author{\bf M. C. Ruivo}
%\email{maria@teor.fis.uc.pt}
\author{\bf C. A. de Sousa}
%\email{celia@teor.fis.uc.pt}
\affiliation{Departamento de F\'{\i}sica, Universidade de
Coimbra, P-3004-516 Coimbra, Portugal}
\author{\bf H. Hansen}
%\email{hansen@ipnl.in2p3.fr}
\affiliation{Univ.Lyon/UCBL, CNRS/IN2P3, IPNL ``Labo en lutte'' , 69622
Villeurbanne Cedex, France}
\author{\bf W.M. Alberico}
%\email{alberico@to.infn.it}
\affiliation{Dipartimento di Fisica Teorica, University of
Torino and INFN, Sezione di Torino, via P. Giuria 1, I-10125 Torino, Italy}
\date{\today}

\begin{abstract}
The modification of mesonic observables in a hot medium is analyzed as a tool to
investigate the restoration of chiral and axial symmetries in the context of the
Polyakov-loop extended Nambu--Jona-Lasinio model. The results of the extended model lead
to the conclusion that the effects of the Polyakov loop are fundamental for reproducing
lattice findings. In particular, the partial restoration of the chiral symmetry is faster
in the PNJL model than in the NJL one, and it is responsible for several effects: the
meson-quark coupling constants show a remarkable difference in both models, there is a
faster tendency to recover the Okubo-Zweig-Iizuka rule, and finally the topological susceptibility
nicely reproduces the lattice results around $T/T_c\approx 1.0$.
\end{abstract}


\maketitle

%%%%%%%%%%%%%%%%%%%%%%%%%%%%%%%%%%%%%%%%%%%%%%%%%%%%%%%%%%%%%%%%%%%%%%%%
%%%%%%%%%%%%%%%%%%%%%%%%%%%%%%%%%%%%%%%%%%%%%%%%%%%%%%%%%%%%%%%%%%%%%%%%

\section{Introduction}

In recent years several studies have been developed, which are concerned with the
properties of matter under extreme conditions of density and/or temperature: 
the restoration of symmetries (e.g., the chiral symmetry) and the phenomenon of
deconfinement, which might be achieved in ultrarelativistic heavy-ion collisions or in
the interior of neutron stars, deserves special attention. Properties of hadrons, in particular,
of mesons, propagating in a hot or dense medium can shed light on the occurrence of the expected
phenomena~\cite{datta1,shury2,wong1,albe,mocsy}. For example, a criterion to
identify an effective restoration of chiral (axial) symmetry is to look for
the degeneracy of the respective chiral partners \cite{costa:PRD}.

The study of meson properties, in the SU$_f$(2) sector, around
the critical region where the phase transition takes place, was performed in
Ref.~\cite{Hansen:2006PRD}, in the framework of the modified Nambu--Jona-Lasinio model
including the Polyakov loop (the so-called PNJL
model)~\cite{Meisinger:1995ih,Meisinger:2001cq,Pisa1,Fukushima:2003fw,
Mocsy:2003qw,Megias:2006PRD,Ratti:2005jh,Robner,Sasaki,mus2,Schaefer:2007PRD}. In the
PNJL model, quarks are simultaneously coupled to the chiral condensate and to the
Polyakov loop, so the model incorporates features of both chiral and $\Z_{N_c}$ symmetry
breaking \cite{Hansen:2006PRD}. The coupling to the Polyakov loop is fundamental for
reproducing lattice results on QCD thermodynamical quantities~\cite{Ratti:2005jh}, since
it produces a suppression of the unphysical colored states (one or two quark states),
which should not contribute to the thermodynamics below the critical temperature.

In this paper, we intend to extend the investigation of light scalar and pseudoscalar
mesons at finite temperature, generalizing previous works to the  SU$_f$(3) sector. In
particular, it will be interesting to compare the properties (e.g., the masses) of the
scalar mesons ($\sigma,\,f_0,\,a_0$, and $K^*_0$) with those of the pseudoscalar nonet
($\eta,\eta^\prime,\pi^0$, and $K$), which can be considered as chiral partners of the
former. We focus our attention on the role of the Polyakov loop in determining the
behavior with the temperature of these mesons.

An interesting open question we wish to address is whether  both  chiral
SU$(N_f)\otimes$SU$(N_f)$ and axial U$_A$(1) symmetries are restored and which
observables could carry information about these restorations. Moreover, we shall
investigate the role of the U$_A$(1) anomaly which, as  is well known, is responsible for
the flavor mixing effect that  removes the degeneracy among several mesons. The U$_A$(1)
symmetry does not exist at the quantum level being explicitly broken  by the axial
anomaly~\cite{Weinberg} which, in turn, can be described at the semiclassical level by
instantons~\cite{t Hooft}.

The flavor mixing induced by the presence of the axial anomaly causes a violation of the
Okubo-Zweig-Iizuka rule, both for scalar and pseudoscalar mesons, hence the restoration  of axial
symmetry should have relevant consequences on the phenomenology of the mesonic mixing
angles as well as on the topological susceptibility. Also in this context the addition of
the Polyakov loop might influence the tendency toward the recovery of the pseudoscalar
and scalar mixing angles, already evaluated within the pure NJL model.

The paper is organized as follows: in Sec. II, we present the model Lagrangian of PNJL
in SU$_f$(3) including the 't Hooft interaction term; specific subsections are dedicated
to the gap equations, and to the properties (masses and mixing angles) of the
pseudoscalar and scalar meson nonets. In Sec. III, we show our results, starting with a
discussion on the characteristic temperatures and the role played by the strange
components; then we display the mesonic masses, the meson-quark coupling constants, and
the topological susceptibility. Our conclusions are reported in Sec. IV.

%%%%%%%%%%%%%%%%%%%%%%%%%%%%%%%%%%%%%%%%%%%%%%%%%%%%%%%%%%%%%%%%%%%%%%%%
%%%%%%%%%%%%%%%%%%%%%%%%%%%%%%%%%%%%%%%%%%%%%%%%%%%%%%%%%%%%%%%%%%%%%%%%

\section{Model and formalism}

%%%%%%%%%%%%%%%%%%%%%%%%%%%%%%%%%%%%%%%%%%%%%%%%%%%%%%%%%%%%%%%%%%%%%%%%

\subsection{The PNJL model with three flavors}

We perform our calculations in the framework of an extended  SU$_f$(3)
PNJL Lagrangian, which includes the 't Hooft instanton induced
interaction term that breaks the U$_A$(1) symmetry; moreover
quarks are coupled to a (spatially constant) temporal background gauge field
representing the Polyakov loop \cite{Fu:2007PRD,Ciminale:2007,Fukushima:2008PRD}
\begin{eqnarray}
{\mathcal L_{PNJL}\,}&=& \bar q(i \gamma^\mu D_\mu-\hat m)q + 
\frac{1}{2}\,g_S\,\,\sum_{a=0}^8\, [\,{(\,\bar q\,\lambda^a\, q\,)}
^2\,\,+\,\,{(\,\bar q \,i\,\gamma_5\,\lambda^a\, q\,)}^2\,] \nonumber\\
&+& g_D\,\{\mbox{det}\,[\bar q\,(1+\gamma_5)\,q] +\mbox{det}
\,[\bar q\,(1-\gamma_5)\,q]\} \nonumber\\
&-& \mathcal{U}\left(\Phi[A],\bar\Phi[A];T\right). \label{eq:lag}
\end{eqnarray}
Here, $q = (u,d,s)$ is the quark field with three flavors ($N_f=3$) and three colors
($N_c=3$), $\hat{m}=\mbox{diag}(m_u,m_d,m_s)$ is the current quark mass matrix, and
$\lambda^a$ are the flavor SU$_f$(3) Gell-Mann matrices ($a = 0,1,\ldots , 8$), with ${
\lambda^0=\sqrt{\frac{2}{3}} \,  {\bf I}}$. The covariant derivative is defined as
$D^{\mu}=\partial^\mu-i A^\mu$, with $A^\mu=\delta^{\mu}_{0}A^0$ (Polyakov gauge); in
Euclidean notation $A^0 = -iA_4$.  The strong coupling constant $G_{Strong}$ is absorbed in the
definition of $A^\mu(x) = G_{Strong} {\cal A}^\mu_a(x)\frac{\lambda_a}{2}$, where ${\cal
A}^\mu_a$ is the (SU$_c$(3)) gauge field and $\lambda_a$ are the (color) Gell-Mann matrices.

The Polyakov loop field  $\Phi$ appearing in the potential term of
(\ref{eq:lag}) is related to the gauge field through the gauge covariant
average of the Polyakov line~\cite{Pisa1,Ratti:2005jh}
\begin{equation}
\Phi(\vec x)=\left\langle \left\langle l(\vec x)\right\rangle\right\rangle
=\frac{1}{N_c}{\rm Tr}_c\left\langle \left\langle L(\vec x)\right\rangle\right\rangle,
\label{eq:phi}
\end{equation}
where
\begin{equation}
L(\vec x) ={\cal P}\exp\left[i\int_0^\beta d\tau A_4(\vec x, \tau)\right]\,.
\label{eq:loop}
\end{equation}
The Polyakov loop is an order parameter for the restoration of the ${\mathbb Z}_3$ (the
center of SU$_c$(3)) symmetry of QCD and  is related to the deconfinement phase
transition: ${\mathbb Z}_3$ is broken in the deconfined phase ($\Phi \rightarrow 1$) and
restored in the confined one ($\Phi \rightarrow 0$) \cite{Polyakov,tHooftNPB,Svetitsky}.

Here, it is important to make some remarks about the applicability of the PNJL model.
Beyond the chiral pointlike coupling between quarks, in the PNJL model, the gluon
dynamics is reduced to a simple static background field representing the Polyakov loop
(see details in Refs.~\cite{Ratti:2005jh,Hansen:2006PRD}). This scenario cannot be
expected to work outside a limited range of temperatures. Indeed, at large temperatures
it is expected that transverse gluons start to be thermodynamically active degrees of
freedom, but   they are not taken into account in the PNJL model. Since, as concluded in
Ref. \cite{Meisinger:2003id}, transverse gluons start to contribute significantly for
$T>2.5\,T_c$, where $T_c$ is the deconfinement temperature, we can assume that the range
of applicability of the model is roughly limited to $T\leq (2-3)T_c$.

Concerning the effective potential for the (complex) $\Phi$ field, there exist in
the literature different choices~\cite{Ratti:2005jh,Robner,Fukushima:2008PRD}: we decided
to adopt the one proposed in Ref.~\cite{Robner} [see Eq. \ref{Ueff}], which is known to give
sensible results \cite{Robner,Sasaki}. In particular, it reproduces, at the mean field
level, results obtained in lattice calculations. The potential reads

\begin{equation}
    \frac{\mathcal{U}\left(\Phi,\bar\Phi;T\right)}{T^4}
    =-\frac{a\left(T\right)}{2}\bar\Phi \Phi +
    b(T)\mbox{ln}[1-6\bar\Phi \Phi+4(\bar\Phi^3+ \Phi^3)-3(\bar\Phi \Phi)^2]
    \label{Ueff}
\end{equation}
where
\begin{equation} a\left(T\right)=a_0+a_1\left(\frac{T_0}{T}\right)+a_2\left(\frac{T_0}{T}
    \right)^2\,\mbox{ and }\,\,b(T)=b_3\left(\frac{T_0}{T}\right)^3.
\end{equation}
%%%%%

We notice that in the mean field approximation the Polyakov loop field $\Phi(\vec{x})$
simply coincides with its expectation value $\Phi=$const., which minimizes the potential
(\ref{Ueff}).  The parameters of the effective potential $\mathcal{U}$ are given in Table
\ref{param}. These parameters have been fixed in order to reproduce the lattice data for
the expectation value of the Polyakov loop and QCD thermodynamics in the pure gauge
sector \cite{latticePL,Kaczmarek:2007}.

The parameter $T_0$ is  the critical temperature for the deconfinement phase
transition within a pure gauge approach: it was fixed to $270$ MeV, according to lattice
findings. Different criteria for fixing $T_0$ are available in the literature, like in
Ref.~\cite{Schaefer:2007PRD}, where an explicit $N_f$ dependence of $T_0$ is presented by
using renormalization group arguments.  Besides, we notice that the Polyakov loop
computed on the lattice with (2+1) flavors and with fairly realistic quark masses is very
similar to the SU$_f$(2) case \cite{Kaczmarek:2007}. Hence, we choose to keep for the
effective potential $\mathcal{U}\left(\Phi,\bar\Phi;T\right)$ the same parameters which
were used in SU$_f$(2) PNJL~\cite{Robner}, including $T_0$.

Since one of the purposes of the present paper is to estimate the effect of the
coupling of the Polyakov loop to quarks with NJL 4-point interactions, we choose to
compare the NJL and PNJL models on a relative temperature scale $T/\Tcc$, where $\Tcc$ is
a characteristic temperature that can be derived in both models (here, the chiral
crossover location). This choice is justified along the lines of the Ginsburg-Landau
effective theory, where characteristic temperatures cannot be absolutely compared between
two models of the same universality class. Besides, as noticed in
Refs.~\cite{Hansen:2006PRD,Schaefer:2007PRD} the $T_0$ dependence of the results is mild.
In the present context the physical outcomes are not dramatically modified when one
changes $T_0$\footnote{The choice of $T_0$ certainly deserves investigations beyond the
scope of this paper.}. Hence the choice $T_0 = 270$ MeV appears to be preferable here,
since it ensures an almost exact coincidence between chiral crossover and deconfinement
at zero chemical potential, as observed in lattice calculations.

With the present choice of the parameters, $\Phi$ and $\bar\Phi$ are always lower than
one in the pure gauge sector. In any case, in the range of applicability of our model
($T\leq 2.5~T_c$), there is a good agreement between our results and the lattice data for
$\Phi$.

%%%%%%%%%%%%%%%%%%%%%%%%%%%%%%%%%%%%
\begin{table}
    \begin{center}
        \begin{tabular}{|c|c|c|c|}
            \hhline{|====|}
            $a_0$ & $a_1$ & $a_2$ & $b_3$ \\
            \hline
            3.51  & -2.47 & 15.2 & -1.75  \\
            \hhline{|====|}
        \end{tabular}
         \caption{ \label{param} Parameters for the effective potential in the pure gauge sector
         (Eq.~(\ref{Ueff})).}
    \end{center}
\end{table}
%%%%%%%%%%%%%%%%%%%%%%%%%%%%%%%%%%%%

Let us anticipate that, at $T=0$, it can be shown   that the minimization of the grand
potential leads to $\Phi=\bar\Phi=0$. So, the quark sector decouples from the gauge one,
and the model is fixed by the coupling constants $g_S, \,g_D$, the cutoff parameter
$\Lambda$, which regularizes the divergent integrals, and the current quark masses $m_i$.
The parameter set used here is
$m_u = m_d = 5.5$ MeV, $m_s = 140.7$ MeV, $g_S \Lambda^2 = 3.67$, $g_D \Lambda^5 -12.36$
and $\Lambda = 602.3$ MeV (for details see Ref.~\cite{costa:2003PRC}).

%%%%%%%%%%%%%%%%%%%%%%%%%%%%%%%%%%%%%%%%%%%%%%%%%%%%%%%%%%%%%%%%%%%%%%%%

\subsection{Gap equations}

From the Lagrangian (\ref{eq:lag}) in the mean field approximation it is straightforward
(see Ref.~\cite{Klevansky-review}) to obtain  effective  quark masses (the gap equations)
given by
%%%%%
\begin{equation}
M_{i}=m_{i}-2g_{_{S}}\left\langle\bar{q_{i}}q_{i}\right\rangle
-2g_{_{D}}\left\langle\bar{q_{j}}q_{j}\right\rangle\left\langle\bar
{q_{k}}q_{k}\right\rangle\,,\label{eq:gap}
\end{equation}
%%%%%
where  the quark condensates $\left\langle\bar{q_{i}}q_{i}\right\rangle$, with
$i,j,k=u,d,s$ (to be fixed in cyclic order), have to be determined in a self-consistent
way. The last term on the right-hand side derives from the determinantal piece in the
Lagrangian, which clearly induces a flavor mixing in the constituent quark masses
$M_{i}$.
So, taking already into account Eq. (\ref{eq:gap}), the PNJL grand canonical  potential
density in the SU$_f$(3) sector can be written as
\be \Omega &=& \Omega(\Phi, \bar\Phi, M_i ; T, \mu) = {\cal
U}\left(\Phi,\bar{\Phi},T\right)
+g_{_{S}}\sum_{\left\{i=u,d,s\right\}}\left\langle\bar{q_{i}}q_{i}\right\rangle^2 +4
g_{_{D}}\left\langle\bar{q_{u}}q_{u}\right\rangle
\left\langle\bar{q_{d}}q_{d}\right\rangle\left\langle\bar{q_{s}}q_{s}\right\rangle \nonumber \\
&& - 2 N_c\sum_{\left\{i=u,d,s\right\}}
\int_\Lambda\frac{\mathrm{d}^3p}{\left(2\pi\right)^3}\,{E_i} - 2
T\sum_{\left\{i=u,d,s\right\}}\int_\Lambda\frac{\mathrm{d}^3p}{\left(2\pi\right)^3}
\left( z^+_\Phi(E_i) + z^-_\Phi(E_i) \right), \label{omega} \ee
where we have defined $ E_i^{(\pm)}\,=\,E_i\,\mp \mu$, the upper
sign applying for fermions and the lower sign for antifermions;
$E_i$ is the quasiparticle energy for the quark $i$:
$E_{i}=\sqrt{\mathbf{p}^{2}+M_{i}^{2}}$; finally, $z^+_\Phi$ and
$z^-_\Phi$ are the partition function densities (with the usual notation $\beta = 1/T$)
\be
z^+_\Phi(E_i) &\equiv& \mathrm{Tr}_c\ln\left[1+ L^\dagger \expp\right]
\ln\left\{ 1 + 3\left( \bar\Phi + \Phi \expp \right) \expp
 + \expppp \right\} \ \label{eq:termo1}\label{zplus}
\\
z^-_\Phi(E_i) &\equiv& \mathrm{Tr}_c\ln\left[1+ L \expm \right]
\ln\left\{ 1 + 3\left( \Phi + \bar\Phi \expm \right) \expm
 + \expmmm \right\}~.\ \label{eq:termo2}\label{zmoins}
\ee

It was shown in Ref.~\cite{Hansen:2006PRD}  that all calculations in the NJL model can be
generalized to the PNJL one by introducing the modified Fermi-Dirac distribution
functions for particles and antiparticles:
%%%%%
\be
f^{(+)}_\Phi(E_i) & = &\frac{ \bar\Phi\expp + 2\Phi\exppp + \expppp }
  {\exp\{z^+_\Phi(E_i)\}} \\
f^{(-)}_\Phi(E_i) & = &\frac{ \Phi\expm + 2\bar\Phi\expmm + \expmmm }
  {\exp\{z^-_\Phi(E_i)\}}. \label{fpPhi}
\ee
%%%%%

To obtain the mean field equations we must search for the minima of the thermodynamical
potential density (\ref{omega}) with respect to
$\left\langle\bar{q}_{i}q_{i}\right\rangle$ ($i=u,d,s$), $\Phi$, and $\bar\Phi$. In fact,
by minimizing $\Omega$ with respect to $\left\langle\bar{q}_{i}q_{i}\right\rangle$, we
obtain the equations for the quark condensates
\begin{equation}
\left\langle\bar{q}_{i}q_{i}\right\rangle\,=\,-\,\,2N_c\int_\Lambda\frac{\mathrm{d}^3p}{\left(2\pi\right)^3}
\frac{M_i}{E_i}[1-f^{(+)}_\Phi(E_i)-f^{(-)}_\Phi(E_i)]\,.
\end{equation}

Furthermore,  minimization of $\Omega$ with respect to $\Phi$ and $\bar\Phi$ provides,
respectively, the two additional mean field equations~\cite{Hansen:2006PRD}

%%%%%
\be
%\frac{\partial\Omega}{\partial\Phi}
0 &=& {T^4} \left\{ -\frac{a(T)}{2}\bar\Phi - 6\frac{b(T)\left[\bar\Phi -2\Phi^2+\bar\Phi^2 \Phi\right]}
{1-6 \bar\Phi \Phi + 4(\bar\Phi^3 + \Phi^3)-3(\bar\Phi \Phi)^2}\right\}
\nonumber\\
   && - 6 T \sum_{\left\{i=u,d,s\right\}}
   \int_\Lambda\frac{\mathrm{d}^3p}{\left(2\pi\right)^3}
   \left( \frac {\exppp}{ \exp\{z^+_\Phi(E_i)\} } + \frac {\expm}{\exp\{ z^-_\Phi(E_i)\} } \right)
\label{eq:domegadfi}  \\
%\frac{\partial\Omega}{\partial\bar\Phi}
0 &=& {T^4} \left\{ -\frac{a(T)}{2}\Phi - 6\frac{b(T)\left[\Phi -2\bar\Phi^2+\bar\Phi \Phi^2\right]}
{1-6 \bar\Phi \Phi + 4(\bar\Phi^3 + \Phi^3)-3(\bar\Phi \Phi)^2}\right\}
\nonumber\\
   && - 6 T \sum_{\left\{i=u,d,s\right\}}
   \int_\Lambda\frac{\mathrm{d}^3p}{\left(2\pi\right)^3}
   \left( \frac {\expp}{ \exp\{z^+_\Phi(E_i)\} } + \frac {\expmm}{\exp\{ z^-_\Phi(E_i)\} } \right)
\label{eq:domegadfib}
\ee
%%%%%
%
This general formalism, presented here for completeness in the grand canonical approach,
is going to be employed in the present work with $\mu=0$. Under this condition,
the constraint $\bar \Phi=\Phi$ holds.

%%%%%%%%%%%%%%%%%%%%%%%%%%%%%%%%%%%%%%%%%%%%%%%%%%%%%%%%%%%%%%%%%%%%%%%%

\subsection{Pseudoscalar and scalar meson nonets}

To calculate the meson mass spectrum, we use the same procedure described
in detail in Ref.~\cite{costa:PRD}.
The model Lagrangian (\ref{eq:lag})  can be put in a form suitable for the usual
bosonization procedure, after reducing the six-quark interaction term in order to obtain
an effective four-quark interaction.  This can be achieved, e.g., by shifting the
operator $(\bar{q} \lambda^a q) \longrightarrow (\bar{q} \lambda^a q) + \left\langle
\bar{q} \lambda^a q\right\rangle$, where $\left\langle \dots \right\rangle$ is the vacuum
expectation value~\cite{Klevansky-review} and by Wick contracting one of the bilinears
$(\bar{q} \lambda^a q)$ in the resulting six-quarks interaction term.
The following effective quark Lagrangian is thus obtained:
\begin{eqnarray}
{\cal L}_{eff} &=& \bar q\,(\,i\, {\gamma}^{\mu}\,\partial_\mu\,-\,\hat m)\, q \,\,
\nonumber \\
&+& \frac{1}{2}S_{ab}[\,(\,\bar q\,\lambda^a\, q\,)(\bar q\,\lambda^b\, q\,)]
+\,\frac{1}{2}P_{ab}[(\,\bar q \,i\,\gamma_5\,\lambda^a\, q\,)\,(\,\bar q
\,i\,\gamma_5\,\lambda^b\, q\,)\,],
\label{lagr_eff}
\end{eqnarray}
where  the projectors $S_{ab}\,, P_{ab}$ are of the form
\begin{eqnarray}
S_{ab} &=& g_S \delta_{ab} + g_D D_{abc}\left\langle \bar{q} \lambda^c
q\right\rangle, \label{sab}\\
P_{ab} &=& g_S \delta_{ab} - g_D D_{abc}\left\langle \bar{q} \lambda^c
q\right\rangle. \label{pab}
\end{eqnarray}
The constants $D_{abc}$ coincide with the SU$_f$(3) structure constants for $a,b,c
=(1,2,\ldots ,8)$, while $D_{0ab}=-\frac{1}{\sqrt{6}}\delta_{ab}$ and
$D_{000}=\sqrt{\frac{2}{3}}$. The bosonization procedure is then realized by integrating
out the quark fields in the functional generator associated with the Lagrangian
(\ref{lagr_eff})  complemented by the coupling to scalar and pseudoscalar bosonic fields.
With this procedure the natural degrees of freedom of low-energy QCD in the mesonic
sector emerge in the resulting effective bosonic action. By expanding the latter up to
second order in the meson fields, one obtains the meson propagators, from which the
masses, meson-quark coupling constants and meson decay constants $f_M$ can be evaluated
according to the usual definitions  (see, for example, Eqs.~(21) and (22) of Ref.~\cite
{Rehberg:1996PRC}).

For example, we obtain the inverse propagator for the pion as
%%%%%
\begin{equation}
D^{-1}_{\pi} (P)= 1-P_{\pi} \Pi^P_{uu} (P),\label{proppion}
\end{equation}
%%%%%
where
%%%%%
\begin{equation}
P_{\pi}=g_{S}+g_{D}\left\langle\bar{q}_{s}q_{s}\right\rangle\,,
\end{equation}
%%%%%
and $\Pi_{ab}^{P}(P)$ is the polarization operator, which in momentum space has the form

%%%%%
\begin{equation}
\Pi_{ab}^{P}(P)=iN_{c}\int\frac{d^{4}p}{(2\pi)^{4}}\mbox{tr}_{D}\left[
S_{i}(p)(\lambda^{a})_{ij}(i\gamma_{5})S_{j}(p+P)(\lambda^{b})_{ji}%
(i\gamma_{5})\right],  \label{actp}
\end{equation}
%%%%%
with $\mbox{tr}_{D}$ being now the trace over Dirac matrices. For details concerning the
calculations for the other meson propagators, see Ref.~\cite{costa:PRD}.  At variance with
Ref.~\cite{costa:PRD}, here the introduction of the Polyakov loop implies, obviously, the
use of the modified Fermi functions  $f_\Phi$ in the calculation.

The inclusion of the 't Hooft interaction in the PNJL/NJL model allows for flavor mixing;
however, by imposing SU$_f$(2) flavor symmetry (namely, $m_u=m_d$) the off-diagonal
coupling strengths that mix the $\pi^0$ with $\eta$ and $\eta'$ vanish; hence, charged
and neutral pions become degenerate in mass, as well as the neutral and charged kaons.

It should be  noticed that flavor mixing somewhat entangles the calculation of the $\eta$ and $\eta'$
masses and couplings, since it gives rise to a $P^2$-dependent mixing angle $\theta_{P}
(P^2)$ \cite {costa:2003PRC,costa:PRD}, which in our scheme is defined as follows:
%%%%%
\begin{equation}
\left(\begin{array}[c]{r}
\eta\\\eta^\prime
\end{array}
\right)=O(\theta_P)\left(\begin{array}[c]{r}
\eta_8\\\eta_0
\end{array}
\right)
=\left(
\begin{array}
[c]{rr}%
\mbox{cos}\theta_{P} & -\mbox{sin}\theta_{P}\\
\mbox{sin}\theta_{P} & \mbox{cos}\theta_{P}%
\end{array}
\right)
\left(\begin{array}[c]{r}
\eta_8\\\eta_0
\end{array}
\right) \,.
\end{equation}
%%%%%
In the above, $\eta$ and $\eta^\prime$ stand for the corresponding physical fields, while
$\eta_8$ and $\eta_0$ are the mathematical objects transforming as  octet and singlet
states of the SU$_f$(3) pseudoscalar meson nonet, respectively. By using a standard
procedure we get the  equation for the mixing angles and for the inverse meson
propagators~\cite{costa:PRD}.

As shown elsewhere, in the framework of the NJL  model \cite{costa:2003PRC,costa:PRD},
the mixing angle $\theta_P$, between the components $\eta_0$ and $\eta_8$ is
$P^2$-dependent; hence, one gets different mixing angles at $P^2=M^2_\eta$ or
$P^2=M^2_{\eta'}$. In the present paper we only discuss the mixing angle at
$P^2=M^2_\eta$; nevertheless, we checked that the behavior of the mixing angle at
$P^2=M^2_{\eta'}$ gives qualitatively similar information.

The same technique used for the  pseudoscalar sector can now be directly applied to the
scalar resonances. We deal here with nine scalar resonances: three $a_0$'s, which are the
scalar partners of the pions, four $\kappa$'s, being the scalar partners of the kaons,
and the $\sigma$ and $f_0$, which are similarly associated with the $\eta$ and
$\eta^{\prime}$. As  in the pseudoscalar case, we impose the  SU$_f$(2) flavor symmetry,
so we have no mixing between the $\sigma$ and $f_0$ and the neutral $a_0^{\,0}$.

As is well known, when the mass of the meson exceeds the sum of the masses of its
constituent quarks, the meson can decay in its quark--antiquark pair, hence becoming a
resonant state.
So, in order to calculate the mass of the resonance, the imaginary part of the propagator
must be taken  into account as well and, following the standard approximation described
in Ref.~\cite{Zhuang}, one can obtain the mass and the corresponding decay width.

%%%%%%%%%%%%%%%%%%%%%%%%%%%%%%%%%%%%%%%%%%%%%%%%%%%%%%%%%%%%%%%%%%%%%%%%
%%%%%%%%%%%%%%%%%%%%%%%%%%%%%%%%%%%%%%%%%%%%%%%%%%%%%%%%%%%%%%%%%%%%%%%%

\section{Numerical results}

%%%%%%%%%%%%%%%%%%%%%%%%%%%%%%%%%%%%%%%%%%%%%%%%%%%%%%%%%%%%%%%%%%%%%%%%

\subsection{NJL vs PNJL: Characteristic temperatures\label{NJLvsPNJLtemperatures}}

\begin{figure}[t]
  \begin{center}
        \hspace{-0.4cm}\includegraphics[width=8.5cm,height=7cm]{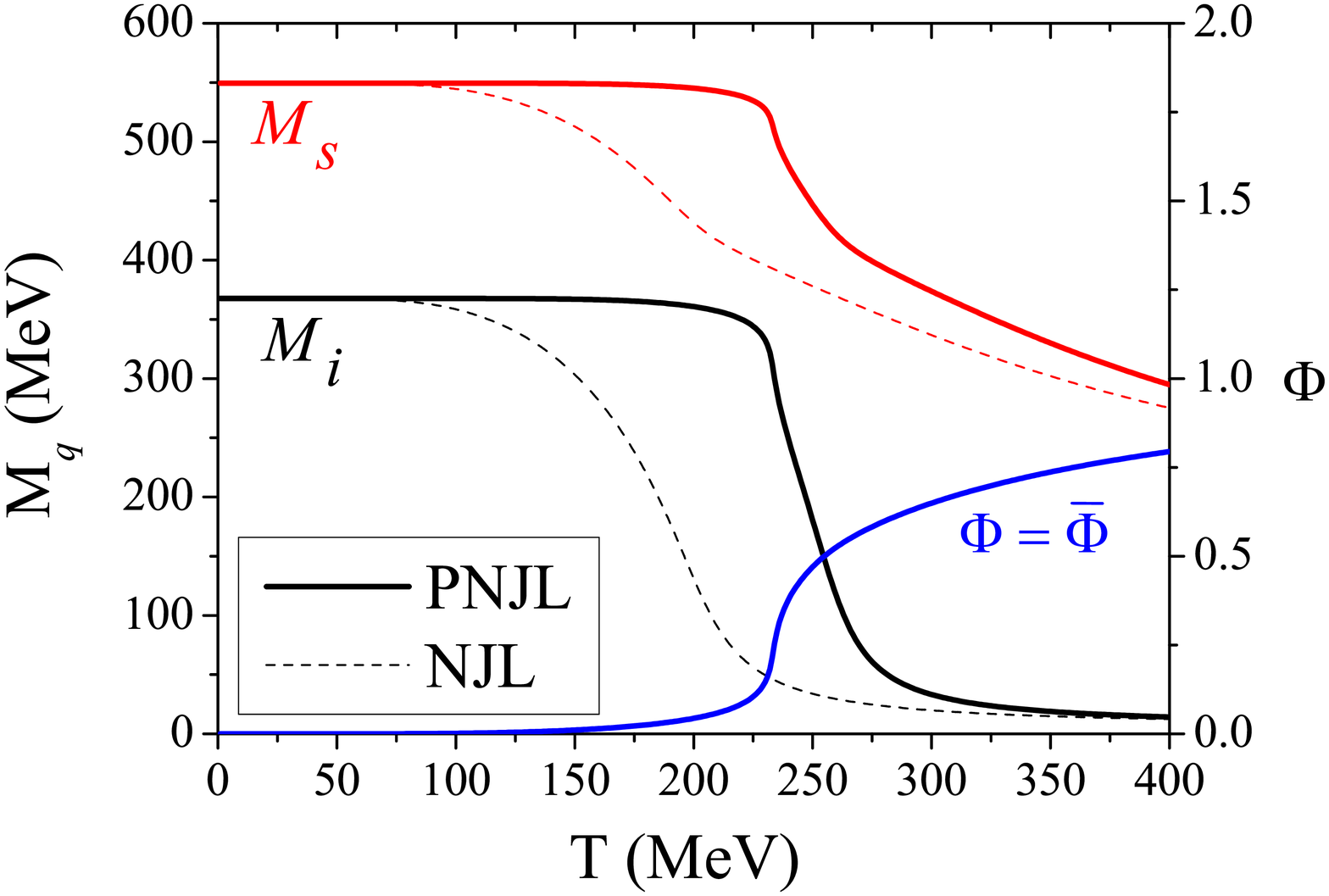}
        \hspace{-0.5cm}\includegraphics[width=8.5cm,height=7.cm]{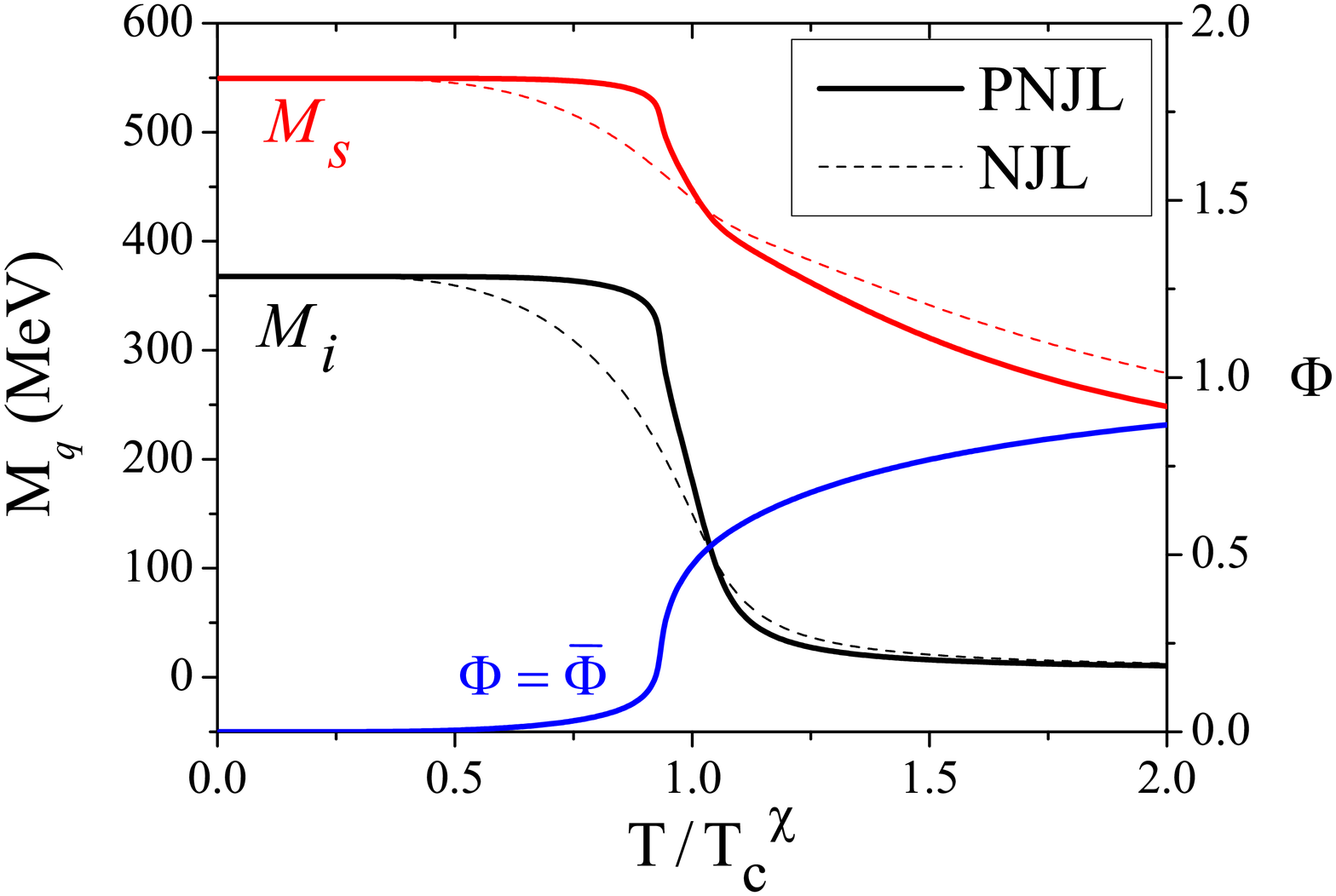}
      \caption{\label{fig:Mquarks} Left panel: comparison of the quark masses
      in the PNJL (solid lines) and NJL (dashed lines) models as functions of the
      temperature; the Polyakov loop crossover is also shown.
      Right panel: the same as before as a function of the reduced
      temperature $T/T_c^{\chi}$.}
  \end{center}
\end{figure}

We start our analysis by identifying the characteristic temperatures that separate the
different thermodynamic phases in PNJL and NJL models \cite{Hansen:2006PRD}. In the
framework of the PNJL model the critical temperature related to the
``deconfinement''\footnote{The terminology ``deconfinement'' in our model is used to
designate the transition between $\Phi \simeq 0$ and $\Phi \simeq 1$ (see
Ref.~\cite{Hansen:2006PRD} for a discussion of this point).} phase transition is
$T_c^\Phi$ and corresponds to the $\Phi$ crossover location (Fig.\ref{fig:Mquarks}). The
chiral phase transition characteristic temperatures, $\Tcc$, are given in both models by
the inflexion points (chiral crossover) of the chiral condensates
$\langle\bar{q_i}q_i\rangle$. Since in both situations chiral symmetry is explicitly
broken by the presence of nonzero current quark mass terms, chiral symmetry is realized
through parity doubling rather than by massless quarks. The effective chiral symmetry
restoration  [in the SU$_f$(2) sector] is signaled by the degeneracy of the chiral
partners $(\pi,\,\sigma)$ and  $(\eta,\,{a_0})$ or, strictly speaking, by the merging of
their spectral functions \cite {Hansen:2006PRD,costa:PRD}.

%%%%%%%%%%%%%%%%%%%%%%%%%%%%%%%%%%%%
%%%%%%%%%%%%%%%%%%%%%%%%%%%%%%%%%%%%

In order to compare the NJL and PNJL results, it is useful to follow the evolution of the
observables as functions of the temperature, expressed both in physical units (MeV) as
well as rescaled in units of a {characteristic} temperature. For the latter we choose the
corresponding $\Tcc$ in NJL and PNJL, and most figures will be shown versus $T/\Tcc$,
which allows a better understanding of the relevant differences between NJL and PNJL models.
Indeed, we are not interested in discussing absolute quantities but rather in comparing
the behavior, below and above $\Tcc$, of the mesonic properties; the key point under
investigation is the effective restoration of symmetries, as well as the influence of the
Polyakov loop on the phase transition.

In Table \ref{temperatures} we quote the {characteristic} temperatures: the effective
chiral symmetry restoration temperature, the deconfinement temperature, and the Mott
temperatures for the pion and the sigma  both in PNJL and NJL models. We remind the
reader that the Mott transition  is associated with the composite nature of the mesons:
at the Mott temperature  the decay of a meson into a $\bar{q}q$ pair becomes
energetically favorable.

As already noticed in the SU$_f$(2) PNJL model~\cite{Ratti:2005jh}, $T_c^\Phi$ differs by
only a few MeV's from $\Tcc$.
In SU$_f$(3) this difference is $17$ MeV. From Table \ref{temperatures}, we also see that
the characteristic temperatures obtained in PNJL are much larger than the lattice result
for the chiral/deconfinement phase transition in (2+1) flavors QCD ($T_c\simeq 170$ MeV).
It was pointed out in Ref.~\cite{Hansen:2006PRD} that the difference between $T_c^\Phi$
and $\Tcc$ is due to the choice of the regularization procedure;  we apply here the
three-dimensional momentum cutoff to both the zero and the finite temperature
contributions. Notice also that a different type of regularization can lower
$\Tcc$~\cite{costa:2007PRD}.
Another important aspect contributing to the present high value of $\Tcc$ is the fact that
we do not rescale the parameter $T_0$ to a smaller value after introducing quarks in the
system. Indeed, we checked that using the lower value $T_0=187$~MeV suggested in
Ref.~\cite{Schaefer:2007PRD},  smaller characteristic temperatures, $T_c^\Phi$ and $\Tcc$
are obtained. However,  we prefer to adopt the higher $T_0$ value since it gives a
smaller difference
between the critical points of the two transitions.
Nevertheless, once we are interested in the general properties of mesons, the
absolute value of the critical temperature is not so relevant: indeed, these properties
are independent of the specific value of $\Tcc$ and a different value of $T_0$ does not
change the conclusions.

%%%%%%%%%%%%%%%%%%%%%%%%%%%%%%%%%%%%
\begin{table}
    \begin{center}
        \begin{tabular}{|c|c|c|c|c|}
        \hhline{|=====|}
                Model &$\Tcc$ [MeV]&$T_c^{\Phi}$ [MeV] &$T_{Mott}^{\pi}$ [MeV]
        &$T_{Mott}^{\sigma}$ [MeV]\\
            \hline PNJL & $ 250$ & $ 233 $
             & $ 267 $ & $ 237 $\\
            \hline NJL  & $196$ & --- &
             $212$ & $160 $\\
        \hhline{|=====|}
        \end{tabular}
    \caption{\label{temperatures} Characteristic temperatures in the NJL and PNJL
    models at $T_0=270$  MeV and zero chemical potential.}
    \end{center}
\end{table}
%%%%%%%%%%%%%%%%%%%%%%%%%%%%%%%%%%%%

%%%%%%%%%%%%%%%%%%%%%%%%%%%%%%%%%%%%%%%%%%%%%%%%%%%%%%%%%%%%%%%%%%%%%%%%

\subsection{The strangeness in PNJL\label{NJLvsPNJLstrangeness}}

In Fig.~\ref{fig:Mquarks} we plot the masses of the strange and nonstrange quarks and
the Polyakov loop as functions of the temperature. At temperatures around $\Tcc$, in both
models, the mass of the light quarks drops to the current quark mass, indicating a smooth
crossover from the chirally broken to an approximate chirally  symmetric phase: we have a
partial restoration of chiral symmetry. This dropping is more pronounced in the PNJL
model than in the NJL one, indicating that the transition toward the partial chirally
restored phase is faster in the former. The strange quark mass shows a behavior very
similar to the one of the nonstrange quarks, with a significant decrease above  $\Tcc$;
however, its mass is still far away from the strange current quark mass. One can say that
chiral symmetry shows a slow tendency to get restored in the strange sector even if this
tendency is faster in the PNJL model.
As in the NJL model, for what concerns the strange sector \cite{costa:PRD} and
since $m_u=m_d<m_s$, the (sub)group SU(2)$\otimes$SU(2) is a much better symmetry of the
Lagrangian (\ref{eq:lag}).

Nevertheless, the fact that the masses of the quarks drop faster around $\Tcc$ in the
PNJL model  is important for the mesonic properties (for example it could modify the
survival of mesonic bound states in the plasma phase).  Moreover, due to the strange
components of some mesons, the behavior of the strange quark mass (modified by the
Polyakov loop) is important for their properties,  as well as for other observables
related to the axial anomaly  (as noticed in \cite{costa:2007PRD} the topological
susceptibility is strongly influenced by the strange sector).

%%%%%%%%%%%%%%%%%%%%%%%%%%%%%%%%%%%%%%%%%%%%%%%%%%%%%%%%%%%%%%%%%%%%%%%%

\subsection{Mesonic masses and mixing angles }

\begin{figure}[t]
\vspace{-0.6cm}
    \begin{center}
            \includegraphics[width=0.64\textwidth]{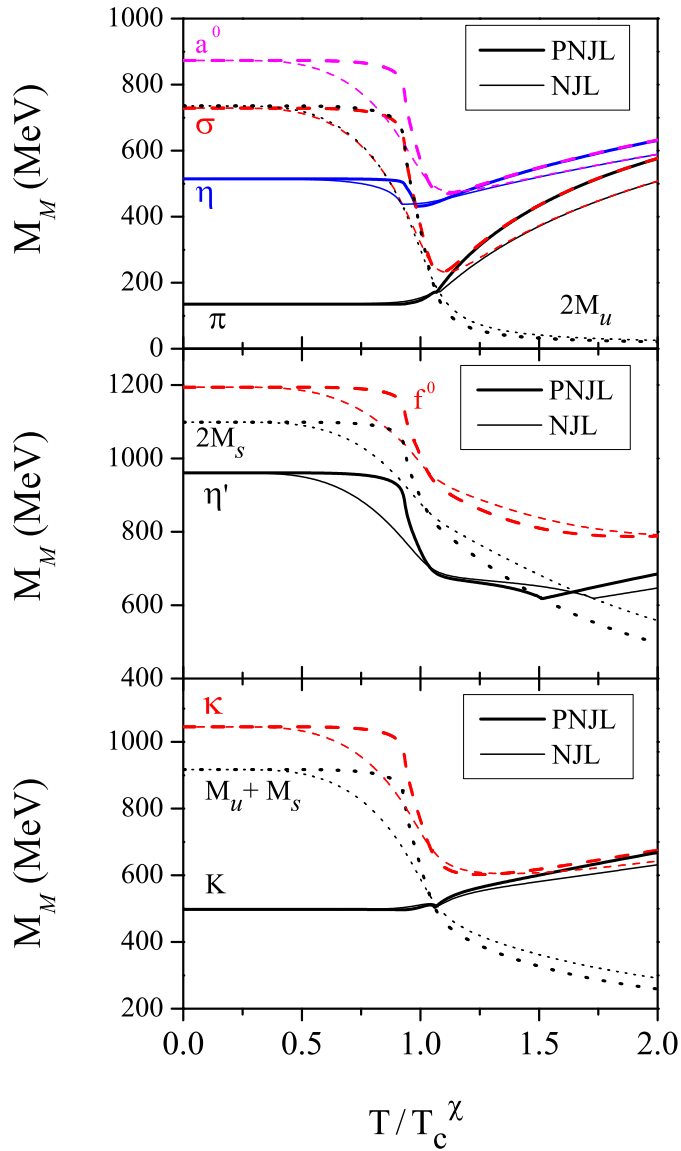}
        \caption{\label{fig:Mmesons} Comparison of the pseudoscalar and scalar mesons
        masses in the PNJL (thick lines) and NJL (thin lines) models as functions
        of the reduced temperature $T/T_c^{\chi}$. In the upper panel
        the $a_0$ (dashed line), $\sigma$ (dotted-dashed line), $\eta$
        and $\pi$ (continuous lines) are shown, together with the
        $2M_u$ mass (dotted lines). In the middle panel $f_0$ (dashed
        line) and $\eta^{\prime}$ (continuous line) are shown and
        compared with the $2M_s$ mass (dotted line). In the lower
        panel the $\kappa$ (dashed line) and $K$ (continuous line)
        masses are compared with $M_u+M_s$ (dotted line). }
   \end{center}
\end{figure}

In Fig.~\ref{fig:Mmesons} we plot the masses of the pseudoscalar mesons and of the respective
scalar chiral partners as functions of the reduced temperature $T/T_c^{\chi}$. The first
evidence emerging from these figures is that the behavior of the mesonic masses in PNJL
looks, qualitatively, very similar to the corresponding one in NJL
\cite{costa:2003PRC,costa:2002PLB}. Hence, the results in the two models will
deserve, in general, similar conclusions, although there are some quantitative
differences with a non-trivial meaning that will be commented on below.

%%%%%%%%%%%%%%%%%%%%%%%%%%%%%%%%%%%%
%%%%%%%%%%%%%%%%%%%%%%%%%%%%%%%%%%%%

In the upper panel of Fig.~\ref{fig:Mmesons} we plot the masses of the scalar and
pseudoscalar mesons  ($\sigma, \,a_{0}, \,\pi, \,\eta$) (upper panel),  $f_0\,,
\eta^{\prime}$ (middle panel) and finally the masses for the $K$ meson and its chiral partner
$\kappa$ (lower panel), both in the NJL and PNJL models. The lower limits of the continua pertaining
to each meson are also shown.
In fact, the continuum starts at the crossing of the $\pi$, $\sigma$, and
$\eta$ lines with the quark threshold $2M_{u}$ (upper panel), of $\eta^\prime$ with
$2M_{s}$ (middle panel) and of the $K$ line with $M_{u}+M_{s}$.

We will now analyze the general behavior of the mesons and mixing angles, starting by
emphasizing what both models have in common.  Concerning the pseudoscalar mesons, it is
found that they are bound states at low temperature (with the exception of the
$\eta'$ meson, which is always above the continuum $\omega_{u}=2M_{u}$),   but at the
respective Mott temperatures they become unbound (see Table \ref{temperatures}); as
usual, for the  $\pi$ and $K$ mesons this occurs at approximately the same temperature in
both models ($T^{K-PNJL(NJL)}_{Mott}= 266~ (210)$ MeV). The $\sigma$  is the only scalar
meson that can be considered as a true (slightly) bound state for small temperatures
(the other mesons being always resonant states) and turns into a resonance at the
corresponding Mott temperature (see Table \ref{temperatures}).

\begin{figure}[t]
\vspace{-0.6cm}
   \begin{center}
        \includegraphics[width=0.64\textwidth]{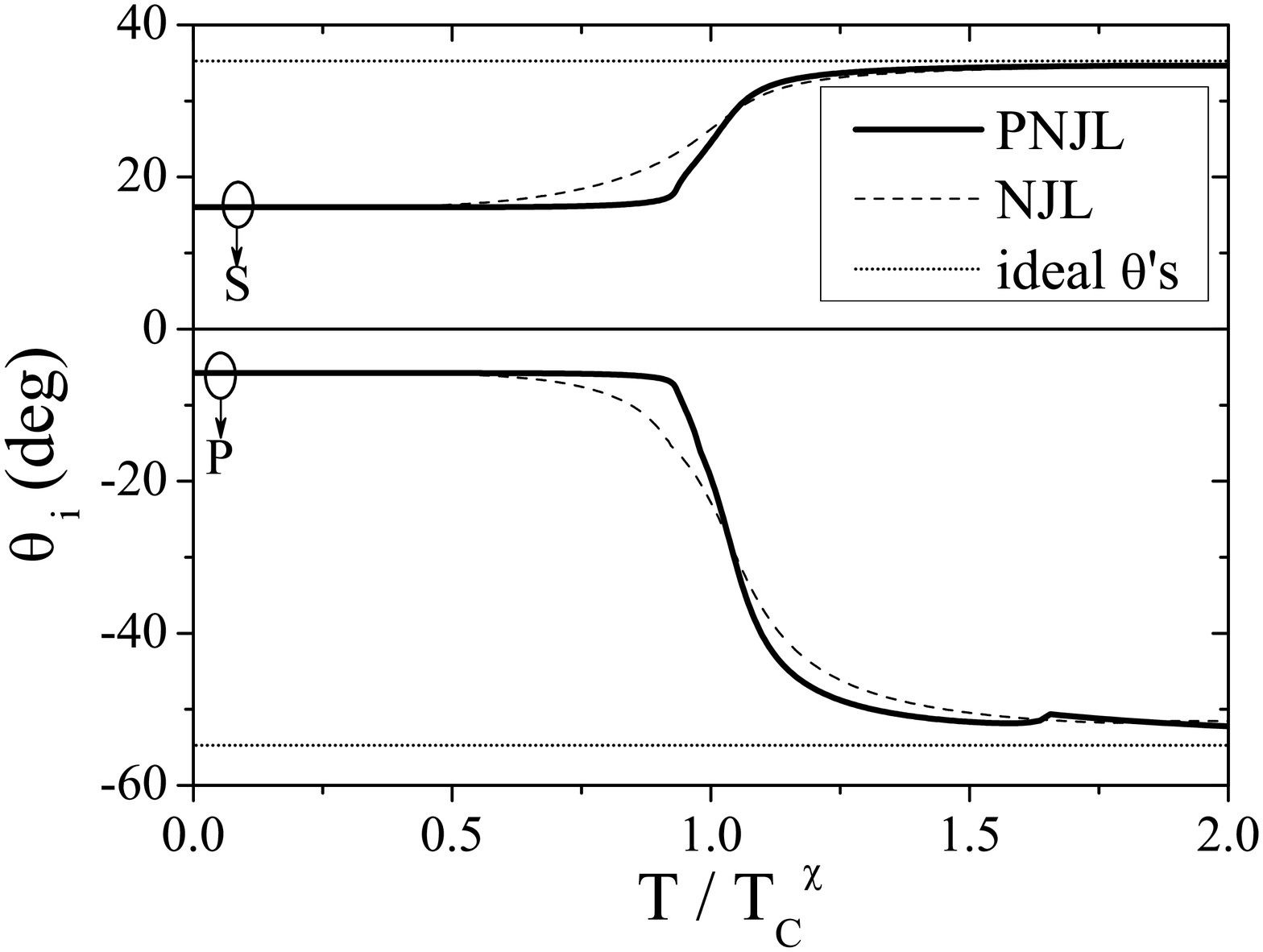}
        \caption{\label{fig:angs} Scalar and pseudoscalar mixing angles as
        a function of the reduced temperature $T/T_c^{\chi}$ for the
        PNJL (solid lines) and NJL (dashed lines) models; the
        ideal mixing angles are also shown (dotted lines).}
    \end{center}
\end{figure}

Concerning the pseudoscalar mixing angle $\theta_{P}$, it is found that, as the
temperature increases, it approaches the ideal value $\theta_{P}=$ $-54.736^{\circ}$,
although never reaching it (see Fig.~\ref{fig:angs}). As a consequence,
the quark content of the  mesons $\eta$ and $\eta'$ changes remarkably,
although a small percentage of mixing always remains: the $\eta$ eventually  becomes
almost nonstrange, while the opposite happens to $\eta'$ \cite{costa:2002PLB,costa:PRD}.
The scalar mixing angle $\theta_S$ exhibits a similar
tendency, the ideal mixing angle $\theta_{S}=$ $35.264^{\circ}$
never being reached in the range of temperatures studied here (see Fig.~\ref{fig:angs});
hence,  the strange component of the $\sigma$ meson decreases but never vanishes,
and $f_0$ becomes almost purely strange.

Our main concern in this paper are the modifications introduced by the Polyakov loop on the
results obtained in the pure NJL model, which are not new (see for instance
\cite{costa:2002PLB,costa:2003PRC,costa:PRD,costa:2007PRD}); yet, some comments are in order
concerning the behavior of the mixing angles, keeping also in mind some recent contributions to
this subject in the framework of other models~\cite{Horvatic:2007,Schaefer:2009}. The
mixing angles are very sensitive to the medium effects, in particular to the influence of the
medium on the strange quark mass:
this might also explain why some aspects of the in-medium  behavior of
$\theta_S$ and $\theta_P$ are not the same in different models or situations.

First of all it should be noted that  the mixing angles depend on the masses of the
mesons and, for the sake of illustration, the angles plotted here are $\theta_S$,
depending on the mass of the $\sigma$ meson and $\theta_P$,  depending on the mass of the
$\eta$ meson. Since the $\sigma$ meson has a small strangeness component, its behavior is
essentially driven  by the decrease of the nonstrange quark mass; on the contrary, the
$\eta$ has a stronger strangeness component, and its behavior is affected not only  by the
fast decrease of the nonstrange quark mass, but also by the slow decrease of the strange
quark mass. Consequently,  the mass of the $\sigma$ decreases more rapidly than the one
of the $\eta$ and, as a result, $\theta_ S$ gets closer than $\theta_P$ to the respective
ideal value, as can be seen in Fig.~\ref{fig:angs}. Notice that there is even a slight
increase of $\theta_P$ at about $T\simeq 1.75 \,\Tcc$, the temperature at which the
$\eta$ meson enters into the strange quark continuum ($m_\eta\geq 2M_s$).

The evolution of the strangeness content of  $\eta,\, \eta'$ determines which one will
become nonstrange, hence behaving as the chiral axial partner of the $\pi$. For example,
the finite temperature behavior of $\theta_P$ leads to the identification of the $\eta$
as the chiral axial partner of the $\pi$ in Refs.~\cite{Schafner:1999, Schafner:2000} and
\cite{costa:2002PLB,costa:2003PRC,costa:PRD,costa:2007PRD}, but  the opposite is found in
Refs.~\cite{Lenaghan:2000,Schaefer:2009}.  An interesting situation was reported in
Refs.~\cite{costa:2002PLB,costa:2003PRC,costa:PRD}, for neutron matter in
$\beta$ equilibrium, where the strange quark mass decreases more rapidly than in
symmetric nuclear matter. As a consequence, the pseudoscalar mixing angle changes sign
and approaches the positive ideal value, while $\eta$ and $\eta'$ exchange identities,
the $\eta' $ becoming nonstrange and exhibiting a tendency to degenerate with the
pion. A level crossing of the pseudoscalar mixing angle and exchange of identities of
$\eta,\, \eta'$ was also found in  Ref.~\cite{Schafner:2000}, and a similar effect for
the scalar angle is reported in  Ref.~\cite{Schaefer:2009}.

Recently, a study of the mixing angles in the framework of a Schwinger-Dyson
approach~\cite{Horvatic:2007}, using a separable interaction, established a connection
between the behavior of  $\theta_P$ and the fastness of restoration of the axial
symmetry, characterized by the drop of the topological susceptibility. Since in the NJL
model the topological susceptibility is proportional to the strange quark condensate,
this result is compatible with our remark, which relates the behavior of the mixing
angles to the strange quark mass evolution. We observe that in the PNJL model there is a
faster restoration of chiral symmetry, both in the nonstrange and strange sectors,
leading to a modification of the in-medium behavior of the mixing angles, meson masses,
coupling constants, and topological susceptibility, as we will discuss below. However, the
modification induced by the Polyakov loop is limited to a range of temperatures around
the critical one ($\sim 0.75\,\Tcc - 1.5\,\Tcc$), and it is not strong enough to alter the
sign of the mixing angles. Having in mind the lack of experimental  information on the
behavior of the observables discussed here, a comparative study of  $\theta_S$ and $
\theta_P$ in different models and situations is desirable.

%%%%%%%%%%%%%%%%%%%%%%%%%%%%%%%%%%%%
%%%%%%%%%%%%%%%%%%%%%%%%%%%%%%%%%%%%%%%%%%%%%%%%%%%%%%%%%%%%%%%%%%%%%%%%

Let us now  analyze the evolution of the mesonic properties in connection with the
possible restoration of symmetries. In the present work, the mesonic behavior is driven
only by the degree of restoration of chiral symmetry in the different sectors. This does
not exclude that other effects, not taken into account here, might influence its
behavior. It can be seen in Fig. \ref{fig:Mmesons} (upper panel) that the partners
($\pi,\sigma$) and ($\eta,a_{0}$) become degenerate  at almost the same temperature. In
both models, this behavior is the signal  of the effective restoration of chiral symmetry
in the nonstrange sector.

On the contrary, the $\eta^{\prime}$ and $f_{0}$ masses do not show a tendency to
converge in the region of temperatures studied, a behavior that reflects the reluctance
of chiral symmetry to get restored in the strange sector.
In fact, due to the
behavior of the mixing angles, $f_0$ and $\eta'$ become essentially strange as the
temperature increases. Moreover, as it has been shown in Ref.~\cite{costa:2007PRD}, even
when the dynamically broken chiral symmetry is restored in all sectors (and differently
from what is found for nonstrange chiral partners) a sizable difference between the
masses of these mesons survives, a fact due to the high value of the current strange
quark mass used here ($m_s=140.7$ MeV). Indeed, at high temperatures  $m_{f_0}^2\simeq
m_{\eta'}^2 + 4 m_s^2$, thus explaining the observed behavior.
Finally, we focus on  the
$ \kappa$ meson (Fig. \ref{fig:Mmesons} lower panel): it is always an unbound, resonant
state and, as the temperature increases, it tends to become degenerate in mass with the
$K$ meson, but at a temperature  of the order of 1.5~$\Tcc$ (in PNJL, and higher in NJL).
In summary, the masses of the mesons that become less strange, $\sigma$ and $\eta$,
converge, respectively, with those of the non strange, $\pi^0$ and $a_0$. The convergence
of the chiral partners $\kappa$ and $K$, which have a $\bar u s$ structure,  occurs  at
higher temperatures,  and is probably slowed down by the  small decrease of the strange
quark mass, $M_s$.

Concerning the axial symmetry, its effective restoration should be signaled by the
vanishing of the observables related to the anomaly, like the mixing angles, the gap
between the masses of the  chiral partners of the U$_A$(1) symmetry, and the topological
susceptibility. For  the observables so far analyzed, we notice that although in both
models the anomaly effects exhibit a tendency to decrease, a full restoration of the
axial symmetry is not achieved: the masses of the partners ($\pi, \eta$) and ($\sigma,
a_0$), although getting close at high temperatures, do not converge and the mixing angles
never reach the ideal values. This was indeed expected since, in the framework of the NJL
model, it has been shown that only with additional assumptions  (for example,  by
choosing a temperature dependent anomaly coefficient~\cite{costa:PRD} or by using a
regularization where the cutoff goes to infinity at $T\not=0$ \cite{costa:2007PRD}) the
restoration of the axial symmetry can be achieved.

Let us now comment on the differences  between the results of
the two models. The new feature of the PNJL model is that the faster decrease of the
quark  condensates  leads to a faster partial restoration of chiral symmetry.  Also the
analysis of the mesonic masses shows that a faster effective restoration of this
symmetry, in the nonstrange sector, is achieved, as can be seen in Fig.
\ref{fig:Mmesons}. In fact, in the NJL model the effective chiral symmetry restoration
for the nonstrange sector occurs at $T_{eff}=1.3~\Tcc$ while, in the PNJL model,
$T_{eff}=1.2~\Tcc$ (again we do a relative comparison between the two models: $T_{eff}$
and $\Tcc$ are derived and compared for each model, respectively);  for the $K-\kappa$
sector, the temperatures are about $1.5$ (NJL) and $2$ (NJL) times the corresponding
characteristic temperature.
Finally, although the axial chiral partners do  not converge, in the PNJL model, the masses
of ($\pi, \eta$) become closer than in NJL, as well as those of ($\sigma, a_0$). From
Fig. \ref{fig:angs} we can also see that around $\Tcc$ the mixing angles $\theta_P$ and
$\theta_S$ approaches faster the ideal angle in the PNJL model than in the NJL one. This
is an indication that, although axial symmetry is not restored in the range of
temperatures studied, the tendency to restore this symmetry is slower in the NJL model.

%%%%%%%%%%%%%%%%%%%%%%%%%%%%%%%%%%%%%%%%%%%%%%%%%%%%%%%%%%%%%%%%%%%%%%%%

\subsection{Coupling constants}

\begin{figure}[t]
\vspace{-0.6cm}
   \begin{center}
        \begin{tabular}[c]{cc}
           \hspace{-0.5cm}\includegraphics[width=7.5cm,height=6.5cm]{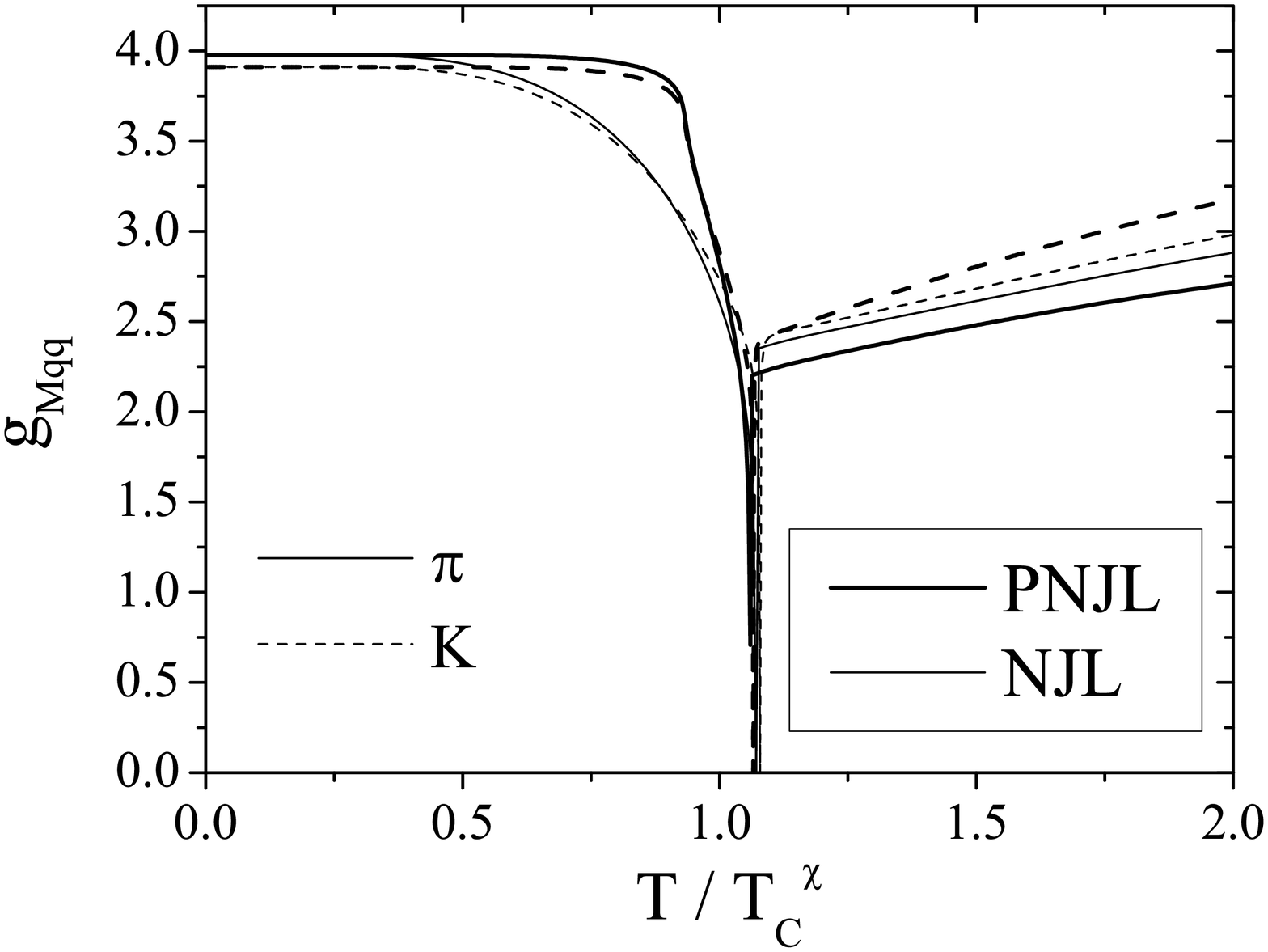}&
            \hspace{-0.5cm}\includegraphics[width=9.5cm,height=7cm]{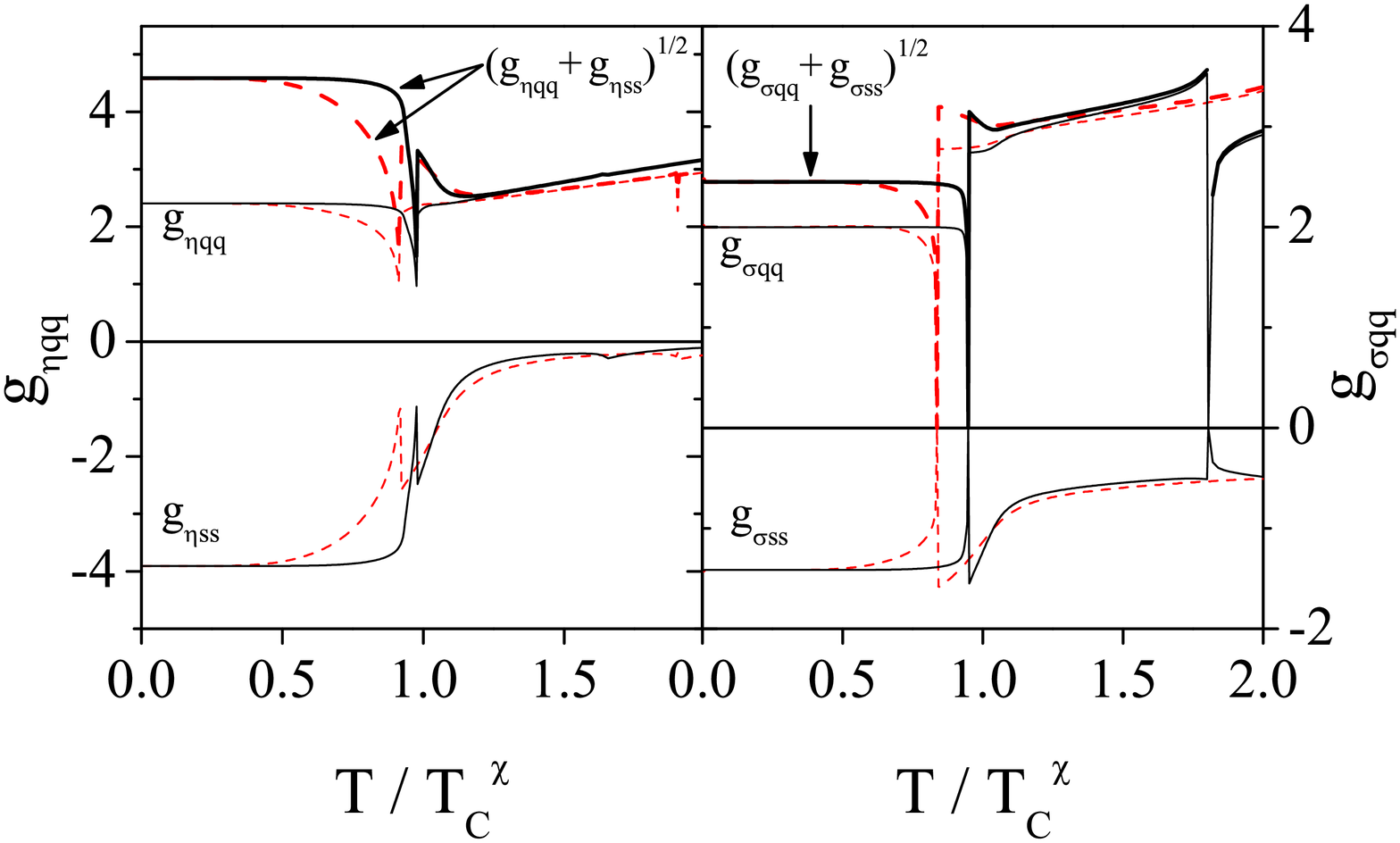}\\
        \end{tabular}
        \caption{\label{fig:acops} The $\pi$, $K$ (left panel) and $\eta$,
        $\sigma$ (right panel) coupling constants as functions
        of the reduced temperature $T/T_c^{\chi}$ for the PNJL (solid lines) and NJL
        (dashed lines) models. }
    \end{center}
\end{figure}

In Fig.\ref{fig:acops} we plot the values of the $\pi$, $K$, $\eta$ and $\sigma$ coupling
constants. We observe a striking behavior at the Mott temperature for each meson: the
coupling strengths approach zero for $T\rightarrow T^{Meson}_{Mott}$
\cite{Rehberg:1996PRC}. This is due to the  fact that the polarization displays a kink
singularity, which can also be seen in the meson masses. For the $\eta$ and $\sigma$
coupling constants there is a second drop toward zero when the mass of these mesons
approach $\omega_s=2M_s$. In other terms, these two zeros signal the entrance into
the continuum for $u,\,d$ quarks and s quarks, respectively.

As already stated, the most striking difference between the NJL and PNJL models lies in the
faster variation with the temperature of the PNJL results around any characteristic temperature.
In particular, close to the phase transition, the NJL and PNJL calculations for
the meson-quark coupling constants show a remarkable difference.
We observe that, in both models, the mesons without flavor mixing ($\pi$ and $K$)
have a higher $T_{Mott}^{Meson}/\Tcc$ ratio than those with flavor mixing ($\eta$ and $\sigma$)
as it can be seen in Table \ref{tempMott}.
It is interesting to note that, while $T_{Mott}^{Meson}/\Tcc$ for $\pi$ and $K$ does not
change appreciably from one model to the other, for $\eta$ and $\sigma$ this ratio
is higher in the PNJL model, where a faster decrease of the mixing effects is observed.
This effect, which indicates a slightly longer survival of these mesons as bound states,
is probably driven by the faster decrease of the strange quark mass, observed in the
PNJL model.

%%%%%%%%%%%%%%%%%%%%%%%%%%%%%%%%%%%%
\begin{table}
    \begin{center}
        \begin{tabular}{|c|c|c|c|c|}
            \hhline{|=====|}
                        &$T_{Mott}^{\pi}/\Tcc$&$T_{Mott}^{\sigma}/\Tcc$&$T_{Mott}^{\eta}/\Tcc$&$T_{Mott}^{K}/\Tcc$\\
            \hline
            PNJL & $ 1.07 $ & $ 0.95 $ & $ 0.98 $ & $ 1.06 $ \\
            \hline
            NJL  & $ 1.08 $ & $ 0.82 $ & $ 0.92 $ & $ 1.07 $ \\
            \hhline{|=====|}
        \end{tabular}
  \caption{\label{tempMott} Reduced Mott temperatures in the NJL and PNJL
  models at zero chemical potential.}
    \end{center}
\end{table}
%%%%%%%%%%%%%%%%%%%%%%%%%%%%%%%%%%%%

The PNJL model is a quantitative step toward confinement with respect to the NJL quark model
because the $\Phi$ factor suppresses the 1- and 2- quarks Boltzmann factor at low
temperature. The fast restoration of the $\Z_3$ symmetry ($\Phi$ goes to one when
temperature increases) producing, in a short range of temperatures, a quark thermal bath
with all (1-, 2- and 3-) quark contributions might explain the fastening of the
transition.

%%%%%%%%%%%%%%%%%%%%%%%%%%%%%%%%%%%%
%%%%%%%%%%%%%%%%%%%%%%%%%%%%%%%%%%%%

%%%%%%%%%%%%%%%%%%%%%%%%%%%%%%%%%%%%%%%%%%%%%%%%%%%%%%%%%%%%%%%%%%%%%%%%

\subsection{Topological susceptibility}

We found it interesting to derive the topological susceptibility $\chi$, which, in
pure color SU(3) theory, is related to the $\eta'$ mass through the Witten-Veneziano
formula \cite{Veneziano}
%%%%%
\begin{equation}
 \frac{2N_f}{f_{\pi}^2}\chi = M_{\eta}^2+M_{\eta^{\prime}}^2- 2M_K^2\,.
\end{equation}
%%%%%

This observable, together with the mesonic masses and  mixing angles, is strongly
influenced by the anomaly:  besides the degeneracy of the axial chiral partners and
the recovery of the OZI rule (mixing angles $\rightarrow$ ideal values), the vanishing
of $\chi$ is an indication of the absence of the anomaly and, consequently, of the
effective  restoration of the U$_A$(1) symmetry.
Lattice calculations indicate a strong decrease of the topological
susceptibility with increasing temperature~\cite{lattice,latticeChu2,bartolome}.

%%%%%%%%%%%%%%%%%%%%%%%%%%%%%%%%%%%%
\begin{figure}[t]
\vspace{-0.6cm}
   \begin{center}
    \includegraphics[width=0.75\textwidth]{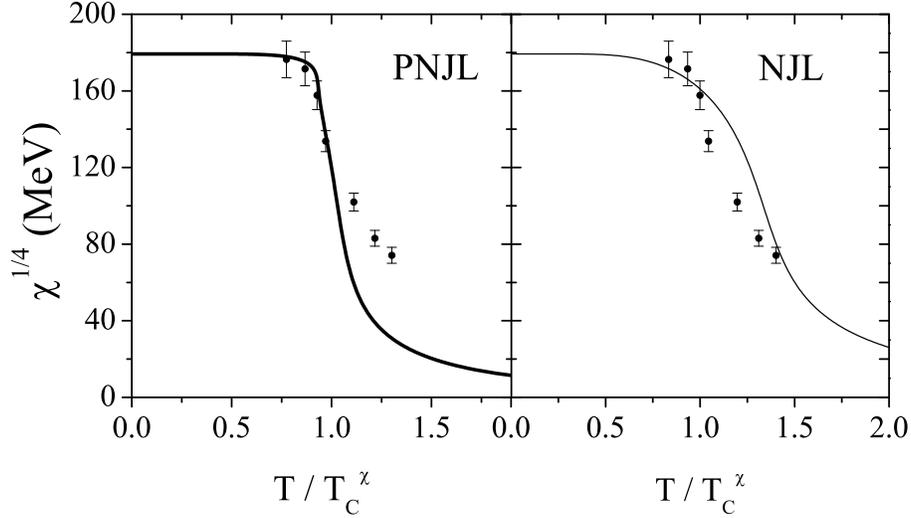}
        \caption{\label{fig:suscep} Topological susceptibility as a function of the reduced
        temperature $T/T_c^{\chi}$ for the PNJL (left panel) and NJL (right panel) models.}
   \end{center}
\end{figure}
%%%%%%%%%%%%%%%%%%%%%%%%%%%%%%%%%%%%

In Fig.~\ref{fig:suscep} we show that, as expected from the previous analysis of the
masses of the chiral partners and mixing angles, the axial symmetry is not fully restored
at high temperatures, and the topological susceptibility is  far away from being zero in
both models (see Fig.~\ref{fig:suscep}). However, at $T=2\Tcc$ the PNJL topological
susceptibility is reduced to about $5\%$ of its value at zero temperature, while the NJL
one has a slower decrease. Moreover, it is interesting to notice that the PNJL calculation
(without any change of the parameters previously fitted, or extra Ansatz, like the
temperature dependence of the anomaly coefficient) nicely reproduces the first lattice
points, namely, the rather steep drop around $\Tcc$, while the NJL model fails to do so.
The faster decrease of  the topological susceptibility, the tendency of the mixing
angles  to approach earlier the ideal values and the reduction of the mass gap of the
axial chiral partners, discussed in Sec. III C,   are  consistent indications that the
Polyakov loop leads to a faster tendency to the restoration of the axial symmetry.

%%%%%%%%%%%%%%%%%%%%%%%%%%%%%%%%%%%%%%%%%%%%%%%%%%%%%%%%%%%%%%%%%%%%%%%%
%%%%%%%%%%%%%%%%%%%%%%%%%%%%%%%%%%%%%%%%%%%%%%%%%%%%%%%%%%%%%%%%%%%%%%%%

\section{Summary and Conclusions}

In this work we have explored the thermodynamical properties of the vacuum state and the
dynamics of the scalar-pseudoscalar meson spectrum propagating in a hot  medium in the
context of a SU$_f$(3) PNJL model. Within the framework of such a model, we have included the
flavor mixing and the coupling of quarks to the Polyakov loop, which in turn is
governed by an effective potential.

Our results indicate that the main feature of the quark masses  is a faster drop around
$T_c^\chi$ in the PNJL model than in the NJL one. This indicates that the partial
restoration of the chiral symmetry is more efficient and fast in the PNJL model. The mass
of the strange quark in the PNJL model is still far from reaching the strange current
quark mass even for high temperatures, although it exhibits a faster decrease than in the
NJL model. This fact contributes to a faster decrease of the anomaly effects, with
implications in the behavior of several observables.

We have observed that, qualitatively, the behavior of mesonic masses in the PNJL model is
similar to the corresponding one in the NJL model. However, we notice  that the mixing angle
approaches the ideal angle faster in the PNJL model.
In addition, the Mott temperatures are different in both models,
showing that  the domain where mesons with flavor mixing are
bound states is extended in the PNJL model.
These results show the relevance of the effects of the interplay among U$_A$(1) anomaly,
the Polyakov loop dynamics, and the partial restoration of the chiral symmetry
at finite temperature.

As a signal  of the effective restoration of chiral symmetry in the nonstrange sector,
the partners ($\pi,\sigma$) and ($\eta,a_{0}$) become degenerate, but this
occurs at a lower reduced temperature in the PNJL model.
On the contrary, in both models the $\eta^{\prime}$ and $f_{0}$ masses do not
show a tendency to converge in the region of temperatures studied, an
indication that chiral symmetry is not likely to be restored
in the  strange sector.

The comparative results of the meson-quark coupling constants are also interesting. In
particular, close to the phase transition, the NJL and PNJL calculations for the
meson-quark coupling constants show meaningful differences. In particular the
$\sigma$ and $\eta$ mesons exhibit a tendency to a slightly longer survival as
bound states.

Finally, there is a significative improvement in the PNJL model
in the results concerning the topological susceptibility.
At $T=2\Tcc$ the latter is reduced to about $5\%$ of its value
at zero temperature, while in the NJL model it exhibits a slower decrease.
Moreover, it is interesting to notice that the PNJL calculation  nicely
reproduces the first lattice points, with a rather steep drop around $\Tcc$,
while this feature is not found in the NJL model.
Although restoration of axial symmetry is not achieved,
this behavior of the topological susceptibility
(and of other relevant observables), indicates that the PNJL model shows a
more efficient mechanism for  the restoration of this symmetry.

%%%%%%%%%%%%%%%%%%%%%%%%%%%%%%%%%%%%%%%%%%%%%%%%%%%%%%%%%%%%%%%%%%%%%%%%

\begin{acknowledgments}
Work supported by Grant No. SFRH/BPD/23252/2005 (P. Costa) and by F.C.T.
under Project Nos. POCI/FP/63945/2005 and POCI/FP/81936/2007 (H. Hansen).
This work was done in spite of the lack of support from Ministry of University and Research
of Italy and France.
\end{acknowledgments}

%%%%%%%%%%%%%%%%%%%%%%%%%%%%%%%%%%%%%%%%%%%%%%%%%%%%%%%%%%%%%%%%%%%%%%%
%%%%%%%%%%%%%%%%%%%%%%%%%%%%%%%%%%%%%%%%%%%%%%%%%%%%%%%%%%%%%%%%%%%%%%%%

\end{document}